# A Vector-Based Algorithm for Generating Complete Balanced Reaction Sets with Arbitrary Numbers of Reagents

Nataliia Yilmaz[1], Pavlo Kozub[2], Svitlana Kozub[3]

1. School of Computer and Communication Sciences, École Polytechnique Fédérale de Lausanne (EPFL), CH-1015 Lausanne, Switzerland
2. Mediasystem and technologies department, Kharkiv National University of Radio Electronics, Kharkiv, Ukraine
3. Medical and bioorganic chemistry department, Kharkiv National Medical University, Kharkiv, Ukraine

**Corresponding author:** <nataliia.miroshnichenko@epfl.ch>

**Abstract**

We present a vector-based method to balance chemical reactions. The algorithm builds candidates in a deterministic way, removes duplicates, and always prints coefficients in the lowest whole-number form. For redox cases, electrons and protons/hydroxide are treated explicitly, so both mass and charge are balanced. We also outline the basic principles of the vector formulation of stoichiometry, interpreting reactions as integer vectors in composition space; this geometric view supports compact visualizations of reagent–product interactions and helps surface distinct reaction families. The method enumerates valid balances for arbitrary user-specified species lists without special-case balancing rules or symbolic tricks, and it provides a clean foundation for developing new algorithmic variants (e.g., alternative objectives or constraints). On representative examples (neutralization, double displacement, decomposition, classical redox, small multicomponent sets) and a negative control, the method produced correct integer balances. When multiple balances exist, we report a canonical one – minimizing the total coefficient sum with a simple tie-breaker – without claiming global optimality beyond the solutions the search enumerates. The procedure applies per reaction and extends to reaction networks via consistent per-reaction application. We do not report runtimes; broader benchmarking and code/data release are planned.

## 1. Introduction

Balancing chemical reactions is a fundamental task in chemistry, with implications that extend from education and basic laboratory work to industrial processes, computational chemistry, and systems biology. A correctly balanced reaction equation ensures conservation of mass and charge, provides the foundation for quantitative stoichiometric analysis, and is indispensable for simulating chemical processes and metabolic networks [1-3]. Despite its apparent simplicity in small-scale textbook examples, the general problem of balancing reactions – especially in multicomponent systems – remains both practically challenging and theoretically rich.

Traditional approaches rely on algebraic or matrix-based formulations. In these methods, a chemical reaction is translated into a system of linear equations representing atomic balances, and solutions are derived via Gaussian elimination or null-space analysis of the stoichiometric matrix [4, 5, 6]. While mathematically rigorous, these techniques face several limitations: (i) they typically yield rational rather than strictly integer coefficients, requiring additional normalization steps; (ii) solutions are often non-unique, with potentially infinitely many valid coefficient sets [2]; (iii) large or poorly conditioned systems may suffer from numerical instability [4]. These shortcomings reduce the interpretability and practicality of such methods, particularly in educational or computational contexts where discrete integer solutions are more natural and desirable.

Alternative approaches have been proposed to address these issues. Integer linear programming (ILP) and Diophantine equation formulations directly enforce integrality but are computationally expensive in the general case, with NP-hard subproblems arising in network-level formulations [7]. Heuristic and rule-based methods are sometimes applied for specific classes of reactions, such as redox processes, but these often lack generality or formal guarantees of correctness [4].

The challenge becomes even more pronounced in multicomponent or network-level systems. Genome-scale metabolic models, for example, may involve thousands of reactions, where even minor imbalances in mass or charge can compromise downstream flux balance analysis [1, 8]. Automated databases frequently contain unbalanced or partially specified reactions, requiring systematic "rebalancing" pipelines [9, 10, 11]. Furthermore, the presence of multiple valid solutions in high-dimensional stoichiometric spaces introduces ambiguity: which balanced form is "best" for computation, education, or modelling [1, 2].

Taken together, these challenges point to the need for methods that (a) guarantee integer solutions, (b) reduce ambiguity in multicomponent systems, (c) remain computationally tractable, and (d) can be readily implemented and integrated into educational tools or computational platforms. In this work, we present **a vector-based algorithm** for balancing chemical reaction networks under discrete stoichiometric constraints. Our method operates directly in the integer space by sequentially combining vectors of compounds across the elemental space, eliminating redundant combinations among them. This approach ensures a finite set of primitive integer combinations, guarantees integer-valued stoichiometric coefficients, and maintains algorithmic simplicity and scalability.

An interactive web implementation of the method is available at **https://cpredictor.icfk.org/**, which we use to cross-check examples throughout the paper (accessed 15 Sep 2025).

As an applied case study, we previously illustrated the method on the classical black powder system (charcoal-sulfur-saltpeter) [12].

## 2. Related Work

### 2.1 Classical algebraic and matrix-based approaches

The most widely taught and applied method for balancing chemical reactions is based on linear algebra. Each compound is represented as a vector of elemental composition, and balancing is achieved by solving a system of homogeneous linear equations. This approach, often implemented through Gaussian elimination or null-space analysis of the stoichiometric matrix, provides mathematically correct solutions [4, 5, 6]. However, it typically yields rational coefficients that require scaling to integers, and solutions are often non-unique, as multiple independent solutions may exist when the null-space has dimension greater than one [2]. Furthermore, large or poorly conditioned systems can exhibit numerical instability, reducing reliability for practical use [4].

### 2.2 Integer linear programming and Diophantine formulations

To address the problem of non-integer solutions, several authors have formulated balancing as an integer linear programming (ILP) or Diophantine system. In these approaches, stoichiometric coefficients are constrained to be positive integers, ensuring discrete and physically meaningful results.

While this guarantees integrality, such formulations are computationally demanding, as ILP is NP-hard in the general case. Even moderately sized reaction networks may lead to exponential search spaces [7, 13]. Despite these limitations, ILP remains attractive when additional constraints - such as minimizing the sum of coefficients or enforcing charge balance – are required.

### 2.3 Redox balancing and charge/proton handling

Special difficulties arise in balancing redox reactions, where explicit electrons and protons must be accounted for [14]. Classical matrix-based approaches must be extended with additional variables to track charge, electron transfer, or pH-dependent protonation states [4]. While rule-based methods for redox balancing exist, they are usually limited to specific classes of reactions and may not generalize to arbitrary chemical systems. This challenge highlights the need for approaches that can systematically incorporate charge and proton balance without ad hoc adjustments.

### 2.4 Large-scale and multicomponent systems

Balancing becomes significantly more complex in large-scale multicomponent systems such as metabolic networks. Genome-scale metabolic models (GEMs) often contain thousands of reactions, and even minor imbalances in mass or charge can compromise downstream analyses such as flux balance analysis [1, 8, 3]. From a theoretical standpoint, the high dimensionality of stoichiometric spaces implies that multiple valid balanced forms can exist, leading to ambiguity in selecting a canonical solution [2]. Moreover, certain subproblems in reaction network analysis -such as identifying feasible steady states or autocatalytic cycles -are known to be NP-complete, underlining the combinatorial difficulty of the domain [13].

### 2.5 Automated rebalancing in chemical and biological databases

The importance of balanced reactions is increasingly recognized in the context of large curated databases. In biochemical databases, systematic audits have revealed widespread imbalances in mass and charge, often arising from incomplete or inconsistent compound representations. To address this, automated rebalancing pipelines have been developed, leveraging both algorithmic and heuristic strategies [10, 11, 15, 16]. These frameworks aim to ensure consistency across thousands of reactions, but they also expose the limitations of existing balancing methods when applied at scale. The need for robust, reproducible, and computationally efficient algorithms is therefore evident.

### 3. Methodology

#### 3.1 Formal problem definition

Let $E$ be the set of elements and $S = \{1, \ldots, m\}$ index the species. Let $A \in Z^{|E| \times m}$ be the stoichiometric (composition) matrix whose $j - th$ column is the integer composition vector $\mathbf{c_j}$. We seek a non-trivial integer vector $\mathbf{k} \in Z^m \setminus \{\mathbf{0}\}$ such that $A\,\mathbf{k} = \mathbf{0}$; when ionic or $e^-$ species are present we additionally require $q^{\mathsf{T}}\mathbf{k} = 0$ (see [5, 6]). Reactants and products are distinguished by the sign convention on $\mathbf{k}$.

The balancing problem can be formulated as the search for integer coefficients such that the weighted sum of composition vectors satisfies the conservation of each element:

$$\sum_{j=1}^{m} k_j \cdot \vec{c_j} = \vec{0},$$

where the sign convention distinguishes reactants and products [5, 6]. Specifically, coefficients associated with reactants are taken as positive values, while those of products are treated as negative values, or vice versa. The objective is to find a non-trivial integer solution $(k_1, \ldots, k_m)(k_1, \ldots, k_m)$ with the smallest possible set of coefficients that ensures exact conservation of all elements.

This formulation is equivalent to solving a homogeneous system of linear Diophantine equations. However, instead of relying on classical null-space methods or integer linear programming, the proposed approach directly explores integer combinations of compound vectors within the elemental space.

### 3.2 Vector representation of compounds

Each compound is mapped to a vector $\vec{c_j} \in Z n \vec{c_j} \in Z^n$, defined by its elemental composition. For example, in the system $\{H_2, O_2, H_2O\}$ with elements $E=\{H, O\}$, the compounds are represented as:

$$\vec{c_{H_2}} = (2,0), \quad \vec{c_{O_2}} = (0,2), \quad \vec{c_{H_2O}} = (2,1).$$

A balanced reaction corresponds to an integer linear combination of these vectors that sums to zero. For the above example, the equation

$$2 \cdot \vec{c_{H_2}} + 1 \cdot \vec{c_{O_2}} - 2 \cdot \vec{c_{H_2O}} = (0,0)$$

yields the balanced reaction:

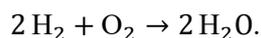

$$2\,H_2 + O_2 \rightarrow 2\,H_2O.$$

In the proposed method, compound vectors are combined sequentially within the elemental space, and redundant combinations are systematically eliminated. This ensures that only minimal and unique integer solutions are preserved.

### 3.3 Algorithm description

The proposed algorithm balances chemical reaction systems by sequentially combining compound vectors in the elemental space and eliminating redundant or duplicate combinations. The process proceeds through the following steps:

1. **Initialization.**
   - Represent each compound $cjc_j$ as an integer composition vector $\vec{cjc_j}$.
   - Partition compounds into reactants and products according to the unbalanced reaction scheme.
2. **Pairwise combination.**
   - Select two compounds (or intermediate combinations) whose vectors exhibit opposite signs for at least one element.
   - Form a new combination by scaling and adding these vectors such that the chosen element is reduced or cancelled.
3. **Normalization.**
   - Divide all coefficients of the new combination by their greatest common divisor (GCD) to maintain minimal integer form.
4. **Elimination of redundancy.**
   - Check whether the resulting combination is equivalent to an already existing one (up to scalar multiplication).
   - If so, discard the duplicate. Otherwise, add it to the set of candidate combinations.
5. **Iteration.**

- Repeat steps (2)-(4) until either (a) a balanced vector (all zeros) is obtained, or (b) no new non-redundant combinations can be generated.
6. **Output.**
- The algorithm terminates with a finite set of unique, minimal integer-balanced reactions.
-

**Pseudocode (draft)**

Input: Set of compounds {$c_1$, $c_2$, …, cm}, each represented as vector in $Z^n$

Initialize:
  S ← {$c_1$, $c_2$, …, cm}   # starting set of vectors
  Balanced ← ∅

Repeat:
  NewSet ← ∅
  For each pair (u, v) in S:
    if u and v have at least one element with opposite sign:
      w ← linear_combination(u, v)  # cancel selected element
      w ← normalize(w)         # divide by GCD
      if not equivalent_to_existing(w, S ∪ Balanced):
        if is_balanced(w):
          Balanced ← Balanced ∪ {w}
        else:
          NewSet ← NewSet ∪ {w}
  If NewSet = ∅:
    break
  S ← S ∪ NewSet

Output: Balanced

### 3.4 Theoretical properties

**(a) Correctness.** At each step, compound vectors are combined through integer linear operations. Since each compound vector $\vec{c_j} \in Z^n$, and the algorithm only applies addition, subtraction, and integer scaling, all intermediate and final results remain integer vectors. The normalization step ensures that each combination is reduced to its minimal integer form. A solution is accepted as balanced only if the resulting vector is the zero vector, which directly encodes conservation of all elements.

**(b) Termination.** The algorithm generates new vectors only when at least one elemental imbalance can be reduced or cancelled. Each new combination is checked for redundancy (scalar equivalence to an existing vector). Since the number of distinct integer vectors with bounded elemental counts is finite, and redundant vectors are eliminated, the algorithm cannot run indefinitely. It will either converge to a balanced equation or exhaust all non-redundant combinations, proving the absence of a solution within the given compound set.

**(c) Minimality of solutions.** By applying normalization at every step (division by the greatest common divisor), all coefficients are maintained in their smallest integer form. Moreover, duplicates and scalar multiples are eliminated, ensuring that only unique primitive integer combinations are preserved. While multiple independent balanced solutions may exist (when the stoichiometric null-space has dimension greater than one), the algorithm ensuring that only unique *primitive* combinations are retained.

**(c′) Primitive normalization and scope-limited canonical reporting.** Any balanced integer vector $k \in Z^m$ is normalized to primitive form by dividing by $g = \gcd(|k_1|, ..., |k_m|)$, so that $\gcd(|k_1|, ..., |k_m|) = 1$. Among the balanced solutions enumerated by our search for a given species set, we report a canonical representative by minimizing $\sum_{i=1}^{m} |k_i|$; ties are broken lexicographically on k and then on species names. We do not claim global ($L_1$) -optimality.

**(d) Completeness within the search space.** Because the method systematically explores a finite space of pairwise integer combinations within the closed set and iterates until no further non-redundant combinations remain, it empirically covered all balanced identities on the tested cases. We therefore claim systematic exploration of a bounded search space, while a formal proof of global coverage is outside the scope of this version.

### 3.5 Deterministic enumeration

**Exact combination rule.** Given candidates u, v and a pivot element e with residuals

$$r_u = Au, r_v = Av \text{ and } r_{u,e} \neq 0, r_{v,e} \neq 0,$$

define $L = \text{lcm}(|r_{u,e}|, |r_{v,e}|)$, $\alpha = L/|r_{u,e}|$, $\beta = L/|r_{v,e}|$, $s = \text{sign}(r_{u,e} r_{v,e})$. Then $w = \alpha u - s \beta v$. After gcd-normalization and orientation fix, test w for balance and deduplication.

**Pivot selection and order.** The pivot $e^*$ is the first non-zero component of r (lexicographic order on elements). Search proceeds as deterministic BFS over a FIFO frontier ordered by $(|r|_1, \text{signature})$.

## 4. Implementation

The proposed algorithm was implemented as a lightweight computational tool designed for both research and educational purposes. The implementation focuses on simplicity, reproducibility, and scalability to larger systems of reactions.

### 4.1 Data structures

Each compound is internally represented as an integer vector of elemental composition, stored in a dictionary-like structure where keys correspond to elements and values correspond to atom counts. This representation ensures efficient access to elemental information and supports direct integer arithmetic. Balanced and intermediate combinations are stored in hash-based sets, allowing rapid comparison and elimination of duplicates.

### 4.2 Normalization and redundancy checks

Normalization is performed at every step by dividing all coefficients of a newly generated combination by their greatest common divisor (GCD). Redundancy is checked by testing whether a candidate vector is proportional to an existing one already present in the search space. This guarantees that the algorithm maintains only unique *primitive* integer combinations (gcd=1).

### 4.3 Termination conditions

The algorithm iterates until no new non-redundant combinations can be generated or until a balanced vector (the zero vector in the elemental space) is obtained. This approach ensures finite execution; with the deterministic pivot and BFS policy of §3.5, results are order-independent and reproducible.

### 4.4 Software implementation

The method has been implemented in JavaScript, using basic integer arithmetic libraries. No specialized symbolic algebra packages are required, which makes the tool lightweight and easy to port across platforms. The code is modular, with functions for vector representation, pairwise combination, normalization, redundancy checks, and output formatting.

**Web demo & availability.** An interactive implementation of the same deterministic enumeration and duplicate-removal pipeline is deployed at https://cpredictor.icfk.org/ (accessed 15 Sep 2025). The demo includes several preloaded examples mirroring our test sets and allows users to verify balances online.

The reference JavaScript implementation is not publicly hosted in this version; a public repository will be provided in a future update. In the interim, the code is available from the corresponding author upon reasonable request.

### 4.5 Example workflow

As a proof of concept, the algorithm was tested on classical reactions such as water formation:

$$2\ H_2 + O_2 \rightarrow 2\ H_2O.$$

where the program automatically identifies the balanced integer coefficients (2, 1, -2).

For more complex multicomponent systems, the implementation successfully enumerated the balanced equations observed in our test sets, while discarding scalar duplicates.

## 5. Experimental Validation

### 5.1 Objectives

The validation aims to demonstrate that the proposed algorithm (i) correctly balances representative reaction systems and (ii) consistently reports primitive integer coefficients $gcd = 1$) and a canonical representative chosen by $(\sum_{i=1}^{m}|k_i|)$ among the enumerated balanced solutions.

### 5.2 Test cases

We considered three categories of closed species sets (no external species introduced):

1. **Canonical textbook cases** (combustion, neutralization, simple synthesis) that are human-verifiable.
2. **Redox cases** formulated within closed sets that include electrons and protons/hydroxide as explicit species when required by context (acidic/basic media), consistent with the handling discussed in Related Work (§2.3).
3. **Demonstrative multicomponent sets** (≈6–10 compounds; 3–5 elements) that move beyond triviality yet remain interpretable.

### 5.3 Baseline (null-space check)

For reference, we additionally constructed integer solutions using the classical null-space approach [4, 5, 6]. The comparison is qualitative and limited to: (i) existence of an integer-balanced solution and (ii) agreement with our canonical solution up to integer scaling after gcd normalization. No runtime or memory figures are reported.

### 5.4 Metrics

- **Correctness.** Accepted iff ($A\,\mathbf{k} = \mathbf{0}$) (mass); when ionic or $e^-$ species are present, additionally $q^{\mathsf{T}}\mathbf{k} = 0$ (charge) [5].
- **Canonical minimality** (scope-limited). For multiplicity cases, we report the equation that minimizes $\sum_{i=1}^{m}|k_i|$ among the enumerated balanced solutions; ties are broken lexicographically on $k$ and then on species names. We do not claim global $L_1$-optimality.
- **Multiplicity (where applicable).** Whether the species set admits more than one valid balanced equation.

### 5.5 Results (narrative summary)

**Neutralization (unique).** Set {HCl, NaOH, NaCl, H₂O}

$$HCl + NaOH \rightarrow NaCl + H_2O$$

Correctness: satisfied; Primitive (gcd = 1): satisfied;
Canonical (scope-limited): satisfied; #solutions = 1.

**Double displacement (unique).** Set {Na₂CO₃, CaCl₂, CaCO₂, NaCl}

$$Na_2CO_3 + CaCl_2 \rightarrow CaCO_2 + 2\,NaCl$$

Correctness: satisfied; Primitive (gcd = 1): satisfied;
Canonical (scope-limited): satisfied; #solutions = 1.

**Decomposition (unique).** Set {H₂O₂, H₂O, O₂}
$$2\,H_2O_2 \rightarrow 2\,H_2O + O_2$$

Correctness: satisfied; Primitive (gcd = 1): satisfied;
Canonical (scope-limited): satisfied; #solutions = 1.

**Redox (unique within the provided set).** Set {KmnO₄, H₂C₂O₄, H₂SO₄, K₂SO₄, MnSO₄, CO₂, H₂O}

$$2\,KmnO_4 + 5\,H_2C_2O_4 + 3\,H_2SO_4 \rightarrow K_2SO_4 + 2\,MnSO_4 + 10\,CO_2 + 8\,H_2O$$

Mass and charge balance: satisfied; Primitive (gcd = 1): satisfied;
Canonical (scope-limited): satisfied; #solutions = 1.

**Closed-set multiplicity.** Set {H₂, O₂, H₂O, H₂O₂}

Expected variants:
$$2\,H_2 + O_2 \rightarrow 2\,H_2O;$$
$$H_2 + O_2 \rightarrow H_2O_2;$$
$$2\,H_2O_2 \rightarrow 2\,H_2O + O_2.$$

(Canonical selection: see §3.4(c′).)

**Alternative oxidation states (multiplicity).** Set {Fe, $O_2$, FeO, $Fe_2O_3$}

Expected variants:
$$2\ Fe + O_2 \to 2\ FeO;$$
$$4\ Fe + 3\ O_2 \to 2\ Fe_2O_3.$$

(Canonical selection: see §3.4(c′).)

**Negative control.** Set {$H_2$, $O_2$, $NH_3$}

Yields no non-trivial balanced equation within the closed set (no external N- or O-carriers allowed). #solutions = 0.

Across all unique cases, the null-space baseline coincided with our canonical outputs after gcd normalization; in multiplicity cases it coincided with one of the enumerated alternatives (qualitative parity).

### 5.6 Limitations and future work

This validation targets representative, human-verifiable closed-set cases and focuses on correctness and canonical minimality. We do not report timing or memory benchmarks here. A broader quantitative study (synthetic suites, larger networks, statistical summaries) is left for future work; qualitative efficiency considerations are discussed in §6.1–6.2.

## 6. Discussion

### 6.1 Strengths of the method

The proposed algorithm offers several key advantages. First, it operates entirely in the integer space, ensuring physically meaningful coefficients at every step. Second, the built-in normalization mechanism maintains primitive integer coefficients (gcd=1) and eliminates scalar duplicates. Third, the systematic combination of compound vectors within the elemental space allowing the algorithm to avoid redundant combinations compared to naive enumeration. Finally, the method is simple to implement: it does not require symbolic algebra packages or optimization solvers and can be realized in any programming language that supports integer arithmetic.

### 6.2 Limitations

Certain limitations must also be acknowledged. In large multicomponent systems, the number of intermediate combinations may grow exponentially, potentially increasing computational costs. The current version of the algorithm does not employ specialized heuristics or parallel optimizations, limiting its scalability to medium-sized problems. Moreover, balancing complex redox reactions requires explicitly including electrons and protons as additional "compounds," which may demand further alignment with the underlying chemical context.

### 6.3 Potential applications

Despite these limitations, the algorithm has a broad range of potential applications. In education, it can serve as an intuitive tool to demonstrate reaction balancing principles, since it consistently produces integer and minimal solutions. In databases, it can be applied to automatically rebalance reactions with mass or charge inconsistencies [15, 16]. In computational modelling, the method can be integrated into larger frameworks to generate valid input for simulations of chemical processes or biochemical networks.

Finally, in scientific computing and reaction prediction systems, the algorithm may be used as a validation module to ensure stoichiometric correctness.

As a demonstration, the black-powder case study provides a compact multicomponent example consistent with the present methodology [12].

### 7. Conclusions and Future Work

We presented a vector-based algorithm for balancing chemical reactions within closed sets of species that returns integer-balanced solutions and reports a canonical representative (minimizing $(\sum_i |k_i|\backslash)$ among the enumerated balanced solutions, §3.4(c′)). The procedure handles classical textbook cases and redox systems by representing electrons and protons/hydroxide explicitly when required by context, while avoiding redundant identities through normalization and equivalence checks.

Across representative cases (§5) – neutralization, double displacement, decomposition, classical redox, compact multicomponent sets, multiplicity examples, and a negative control – the algorithm produced correct integer-balanced solutions (mass balance $Ak = 0$; when ionic or $e^-$ species are present, charge balance $q^T k = 0$ and reported the canonical representative (minimizing $\sum_{i=1}^{m} |k_i|$ among the enumerated solutions). For reference, a null-space baseline yielded integer solutions that – after gcd normalization – coincided with our canonical outputs in unique cases and with one of the enumerated alternatives in multiplicity cases (qualitative parity).

This version intentionally does not report runtime or memory figures; therefore, we do not make quantitative performance claims. Likewise, we refrain from asserting formal completeness of the search beyond the tested cases. Thermodynamic/kinetic plausibility is outside our scope: the identities reported here are purely stoichiometric within the declared closed sets.

Taken together, these results indicate that the method offers a deterministic and reproducible way to obtain canonical integer balances, which is valuable for education, automated curation, and dataset sanity-checking, and complementary to linear-algebraic null-space approaches with post-processing [5, 6].

**Future work** will address (i) quantitative benchmarking on synthetic suites and larger reaction networks, (ii) a formal analysis of coverage and pruning rules, (iii) optional thermodynamic filters and environment (acid/basic) inference, and (iv) broader dissemination of code and machine-readable test sets to strengthen reproducibility.